\documentclass[usenatbib]{mnras}
\usepackage{xcolor, graphicx} 
\usepackage{amsmath}
\usepackage{multicol}
\usepackage{relsize}

\newcommand{\da}{d_\mathrm{A}}
\newcommand{\aV}{\alpha_\mathrm{V}}

\newcommand{\Pt}{P^\mathrm{t}}

\title[Bispectrum Distortions]{Mixing Bispectrum Multipoles Under Geometric Distortions}

\author[Khomeriki \& Saumushia]{
Giorgi Khomeriki$^{1}$\thanks{giorgi.khomeriki@phys.ksu.edu} and
Lado Saumushia$^{1}$\thanks{lado@phys.ksu.edu}\\
$^{1}$ 1228 N.~Martin Luther King Jr.~Drive, Department of Physics, Kansas State University, Manhattan, KS 66506-2601 
}

\begin{document}
\label{firstpage}
\pagerange{\pageref{firstpage}--\pageref{lastpage}}
\maketitle

\begin{abstract}
We derive general expressions for how the Alcock-Paczynski distortions affect the power spectrum and the bispectrum of cosmological fields. We compute explicit formulas for the mixing coefficients of bispectrum multipoles in the linear approximation. The leading-order effect for the bispectrum is the uniform dilation of all three wavevectors. The mixing coefficients depend on the shape of the bispectrum triplet. Our results for the bispectrum multipoles are framed in terms of the "natural" basis of the lengths of three wavevectors but can be easily generalized for other bases and reduction schemes. Our validation tests confirm that the linear approximation is extremely accurate for all power spectrum multipoles. The linear approximation is accurate for the bispectrum monopole but results in sub-percent level inaccuracies for the bispectrum quadrupole and fails for the bispectrum hexadecapole. Our results can be used to simplify the analysis of the bispectrum from galaxy surveys, especially the measurement of the Baryon Acoustic Oscillation peak position. They can be used to replace numeric schemes with exact analytic formulae.
\end{abstract}


\maketitle

\section{Introduction}

Galaxy surveys naturally record the positions of galaxies in spherical coordinates by measuring their angular positions and redshifts. We need to know the curvature of the Universe and its expansion history to convert these to Cartesian coordinates in physical units. If the fiducial model used for this conversion is offset from the real physics of the Universe, the distances between the galaxies will be distorted. These distortions can be used to derive the real cosmology from the fiducial model \citep{1979Natur.281..358A}. This  affects all statistical measures of the distribution of galaxies. For example, the measured power spectrum of galaxies will be a distorted version of the true power spectrum
\begin{equation}
P^\mathrm{m}(k_{||} k_\perp) = P^\mathrm{t}(\alpha_{||} 
\left [ \textbf{p} \right ] k^\mathrm{m}_{||},  \alpha_\perp \left [ \textbf{p} \right ] k^\mathrm{m}_\perp),
\end{equation}
where $\alpha_{||}$ and $\alpha_\perp$ are the geometric distortion parameters along and across the line of sight that depend on cosmological parameters $\mathbf{p}$.
If there is a feature in the power spectrum at a specific known wave number, these distortions can be used to constrain $\textbf{p}$, by finding the values of $\alpha$ parameter that align the wavenumber of that feature to its known value in the measured power spectrum.

The geometric distortions are usually parametrized by $\aV$ and $\alpha_{\epsilon}$, where the $\aV$ parameter describes the average distortion of all the distances and the $\alpha_{\epsilon}$ parameter describes the differential distortion of the line-of-sight distance with respect to the across-the-line-of-sight distances. Any measurement that has a feature at a known scale will be sensitive to $\aV$ and any measurement that is known to be isotropic (or if its angular anisotropy before the distortions is known accurately) will be sensitive to $\alpha_{\epsilon}$ \citep{2003ApJ...594..665B,2009MNRAS.399.1663G, 2011MNRAS.410.1993S,2012MNRAS.419.3223K, 2014JCAP...04..001B, 2014PhRvD..89j3541S, 2015MNRAS.451.1331R, 2016arXiv160302389S, 2017IJMPD..2650055M}.

Baryon Acoustic Oscillation (BAO) feature is the most important distance scale that has been used in this fashion. The BAO is a pattern of overdensities in the initial distribution of matter \citep{2005MNRAS.362..505C,2005ApJ...633..560E,2017MNRAS.470.2617A} . The feature has been measured in the two-point clustering of many galaxy samples and these measurements provide the bust current constrain on the distance redshift relationship \citep{2015MNRAS.449..835R,2017MNRAS.464.1168R,2021MNRAS.500..736B,2021MNRAS.500.3254R,2021MNRAS.500.1201H,2020ApJ...901..153D} .

Other measurements that have been used for the same purpose include wavelet transformed galaxy fields \citep{2022PhRvD.106j3509V}, cosmic void properties \citep{2017ApJ...835..160M,2020MNRAS.499..587E,2022A&A...658A..20H, 2023MNRAS.522.2553P, 2023MNRAS.521.4731V}, void-galaxy clustering \citep{2019PhRvD.100b3504N,2019MNRAS.485.5761C,2020PhRvL.124v1301N,2020MNRAS.491.4554Z,2022MNRAS.509.1871C,2022MNRAS.516.4307W,2023MNRAS.522.2553P,2023MNRAS.521.4731V,2023MNRAS.523.6360W,2023A&A...674A.185M,2023arXiv230205302R}, Lyman $\alpha$ forest \citep{2013MNRAS.436.1674A,2021MNRAS.506.5439C,2023MNRAS.523.3773C} , HI intensity mapping \citep{2008MNRAS.385L..63N, 2009MNRAS.397.1926W,2019MNRAS.484.1007W,2019PhRvD.100l3522B,2020MNRAS.499..587E,2021MNRAS.502.2549S,2021MNRAS.507.1623C,2022MNRAS.516.5454R,2023A&A...674A.185M}. Recent works suggested that the entire galaxy distribution can be used as a probe to measure the distortion parameters \citep{2019A&A...621A..69R,2023RPPh...86g6901M}.

The main objective of our work is to investigate the details of using galaxy bispectrum (three-point correlation function) multipoles for the same purpose. The bispectrum of galaxy density distribution, $B(k_1, k_2, k_3, \umu_1, \phi)$ is a function of five parameters \citep{2000ApJ...544..597S, 2006PhRvD..74b3522S} . It is often expanded in the spherical harmonics, $B_{\ell m}(k_1, k_2, k_3)$, in its last two arguments for convenience \citep{2017MNRAS.467..928G, 2019MNRAS.484..364S,2021JCAP...03..105B,2023arXiv230403643W}. Forecasts suggest that the bispectrum measurements can significantly enhance the cosmological constraints from the galaxy power spectrum \citep{2017MNRAS.467..928G, 2019MNRAS.483.2078Y,2020MNRAS.497.1684S,2022MNRAS.tmp.2216B,2023MNRAS.520.2611T,2023MNRAS.520.2534M}. Recent works attempted to analyze galaxy bispectrum for measuring the BAO position \citep{2017MNRAS.469.1738S,2017MNRAS.468.1070S,2018MNRAS.478.4500P}, and the full anisotropic shape \citep{2004PhRvD..70h3007S,2017MNRAS.465.1757G,2017MNRAS.471.1581B,2018MNRAS.477.1604G,2020JCAP...03..056O,2021MNRAS.501.2862S,2021JCAP...07..008G,2021JCAP...11..038O,2021JCAP...05..015M,2021JCAP...07..008G,2022PhRvD.105d3517P,2022MNRAS.512.4961A,2023JCAP...01..031R,2023MNRAS.520.2534M,2023MNRAS.523.3133S,2023MNRAS.524.1651S} , baryonic physics \citep{2012MNRAS.420.3469P,2015MNRAS.448....9S,2018MNRAS.474.2109S,2023MNRAS.521.1448Y} , and the primordial non-Gaussianity \citep{2007PhRvD..76h3004S,2010MNRAS.406.1014S,2012JCAP...08..036F,2015PhRvD..91l3518B,2021JCAP...05..015M,2021JCAP...04..013M,2023arXiv230504028K,2022PhRvD.106l3525G,2022PhDT........17C,2023arXiv230503070G,2023ApJ...943..178C,2023ApJ...943...64C}. The clustering patterns of fourth and higher order have also been used in the analysis of galaxy samples \citep{1978ApJ...221...19F,2001ApJ...553...14V,2021JCAP...01..015G,2021JCAP...07..008G,2021arXiv210801670P,2022JCAP...09..050G,2023MNRAS.522.5701H,2023JPhA...56F5204C,2023MNRAS.522.5701H,2023JCAP...01..031R}.

The geometric distortions mix the multipoles and this has to be accounted for in the analysis. The head-on solution is to model geometric distortions directly by rescaling the arguments of the measured bispectrum, but this is not computationally efficient. The power spectrum and the bispectrum are usually analyzed by running Markov Chain Monte Carlo (MCMC) over cosmological parameters \citep{2015MNRAS.453.4384H, 2015A&C....12...45Z, 2019PDU....24..260B,2021JCAP...05..057T}. Recomputing bispectrum models for each point in the MCMC chain is computationally expensive. A faster  option is to estimate the distorted bispectrum from precomputed moments and derivatives at a fiducial cosmology using a linear approximation. This is an established methodology for the power spectrum analysis \citep{2008PhRvD..77l3540P} and has also been used for the bispectrum analysis \citep{2021MNRAS.501.2862S}.

After outlining basic formalism and establishing notation in section~\ref{sec:distortions} we provide a review of how the geometric distortions mix different power spectrum multipoles in section~\ref{sec:powerspectrum}. We then derive the mixing coefficients for the bispectrum multipoles in section~\ref{sec:bispectrum}. The validity of linear expansion is tested in section~\ref{sec:validation}. We summarize our main findings in section~\ref{sec:conclusions}.

\section{Alcock-Patczynski Distortions}
\label{sec:distortions}

We define the Alcock-Patczynski distortion parameters as
\begin{align}
    d_{||} &=\alpha_{||}d^\mathrm{t}_{||},\\
    d_{\perp} &= \alpha_{\perp} d^\mathrm{t}_{\perp}.
\end{align}
They are related to the cosmological parameters through
\begin{align}
    \alpha_{||} &= \frac{H^\mathrm{t}(z)}{H^\mathrm{fid}(z)} ,\\
    \alpha_{\perp} &= \frac{\da^\mathrm{fid}(z)}{\da^\mathrm{t}(z)} .
\end{align}
where $H(z)$ and $\da(z)$ are the Hubble parameter and the angular diameter distance at redshift $z$.
These distortions induce dual distortions in the Fourier space
\begin{align}
    k^\mathrm{t}_{||} = \alpha_{||}k_{||},\\
    k^\mathrm{t}_{\perp} = \alpha_\perp k_\perp.
\end{align}
From now on we will work with parameters 
\begin{align}
    \aV &= \sqrt[3]{\alpha_\perp^2\alpha_{||}},    \label{eq:alphav}\\
    \alpha_{\epsilon} &= \sqrt[3]{\frac{\alpha_{||}^2}{\alpha_\perp^2}}.\label{eq:epsilon}
\end{align}
The definition of $\aV$ is standard in the literature. Our definition of $\alpha_{\epsilon}$ differs from the established one ($\epsilon = 1 - \alpha_{||}/\alpha_\perp$). Either $\alpha_\epsilon$ or $\epsilon$ can be used in analysis without a loss of information but we will later show that the definition of Eq.~\eqref{eq:epsilon} has a clearer geometric meaning. To linear order, the two sets of distortion parameters are related as 
\begin{align}
\delta\alpha_V &= \frac{1}{3} \left ( \delta\alpha_\parallel + 2\delta\alpha_\perp \right ),\\
\delta\alpha_{\epsilon} &= \frac{2}{3}(\delta\alpha_\parallel - \delta\alpha_\perp) .
\end{align}

Every wavevector's length and the angle with respect to the z-axis transform as 
\begin{align}
\label{eq:kfull}
\nonumber
    k^\mathrm{t} &= k\sqrt{\umu^2(\alpha_{||}^2 - \alpha_\perp^2) + \alpha_\perp^2} \\
    &= k\aV\sqrt{\umu^2(\alpha_{\epsilon}^2 - 1/\alpha_{\epsilon}) + 1/\alpha_{\epsilon}},\\
    \label{eq:umufull}
    \nonumber
   \umu^t & = \umu\frac{\alpha_{||}}{\sqrt{\umu^2(\alpha_{||}^2 - \alpha_\perp^2) + \alpha_\perp^2}}\\ &
   = \umu\frac{\alpha_{\epsilon}}{\sqrt{\umu^2(\alpha_{\epsilon}^2 - 1/\alpha_{\epsilon}) + 1/\alpha_{\epsilon}}}.
\end{align}
To the linear order in AP parameters, this transformation is,
\begin{align}
    k^\mathrm{t} &= k\left[1 + \delta\alpha_V + L_2(\umu)\delta\alpha_{\epsilon}\right],\label{eq:klinear}\\
   \umu^\mathrm{t} & = \umu\left[1 + \frac{3}{2}\nu^2\delta\alpha_{\epsilon}\right].\label{eq:umulinear}
\end{align}
$\aV$ is the average relative distortion of a wavevector lengths over all possible directions and $\alpha_{\epsilon}$ is the magnitude of the dipole distortion with respect to the x-axis. The angle of the wavevector is only affected by $\alpha_{\epsilon}$. This simple geometric meaning of the new distortion parameters is the motivation behind definitions of Eqs.~\eqref{eq:alphav}-\eqref{eq:epsilon}.

\begin{figure}
    \centering
    \includegraphics[width=\columnwidth]{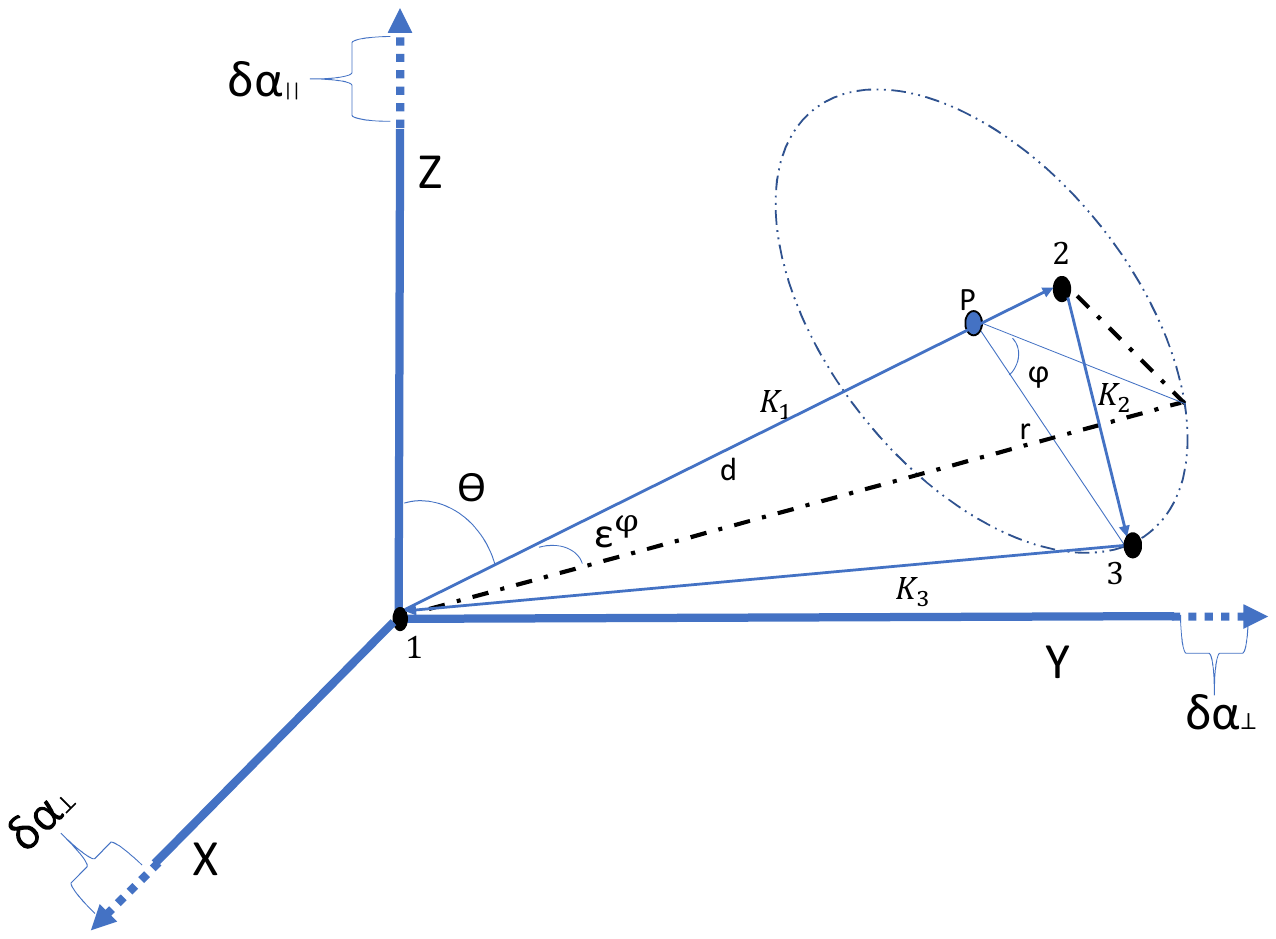}
    \caption{A depiction of three wavevectors defining a bispectrum and associated geometry. Point 3 is in the y-z plane.}
    \label{fig:triangle}
\end{figure}

The bispectrum depends on three wavevectors that satisfy a condition $\mathbf{k}_1 + \mathbf{k}_2 + \mathbf{k}_3 = 0$. These three wavevectors are usually drawn to form a triangle to make this constraint more obvious. The value of the bispectrum only depends on the lengths of the wavevectors $k_1$, $k_2$, $k_3$, and their relative position with respect to the line-of-sight, which is usually fixed by providing two numbers $\umu_1$ and $\phi$. $\umu_1$ is the cosine of the $\mathbf{k}_1$ with respect to the line-of-sight and $\phi$ is the angle of rotation of the triangle from yz plane.

Each of the three wavevectors scales by Eqs.~\eqref{eq:klinear}-\eqref{eq:umulinear}. We need to express the cosine and the sine of these wavevectors in terms of the five basic parameters $k_1$, $k_2$, $k_3$, $\umu_1$, $\phi$ to use these distortion formulae in our computations. These relationships turn out to be relatively simple.
\begin{align}
    \umu_1 &= \umu_\theta,\label{eq:umu1}\\
    \umu_2 &= -\umu_{12}\umu_\theta + \umu_\phi\nu_{12}\nu_\theta,\label{eq:umu2}\\
    \umu_3 &= -\umu_{31}\umu_\theta - \umu_\phi\nu_{31}\nu_\theta,\label{eq:umu3}
\end{align}
where the subscripts 12 and 31 refer to the angles between wavevectors $\mathbf{k}_1$ and $\mathbf{k}_2$, and $\mathbf{k}_3$ and $\mathbf{k}_1$ respectively and
\begin{align}
    \umu_{12} = \frac{\mathbf{k}_1\mathbf{k}_2}{k_1k_2},\\
    \umu_{31} = \frac{\mathbf{k}_3\mathbf{k}_1}{k_3k_1}.
\end{align}
The first relationship is obvious. The cosine of the angle of $\mathbf{k}_1$ is one of the five basic parameters. The other two are derived in Appendix~\ref{app:vectorprojections}. The true and the measured wavevector lengths are related by Eqs.~\eqref{eq:kfull} and \eqref{eq:umufull} for arbitrary distortions and by Eqs.~\eqref{eq:klinear} and \eqref{eq:umulinear} for small distortions, with the angles from Eqs.~\eqref{eq:umu1}-\eqref{eq:umu3}. 

The angle $\phi$ also changes with distortions, but this has a subdominant effect on the multiples and we ignore it in our computations.

\section{Distortions of the Power Spectrum}
\label{sec:powerspectrum}

The power spectrum is a function of a wavenumber length and its angle with respect to the line-of-sight and is usually expanded in Legendre moments,
\begin{equation}
 P (k, \umu) = \sum_l P_l (k) L_l (\umu).
\end{equation}
The moments can be computed as
\begin{equation}
    P_l (k) = \frac{2l+1}{2} \int \mathrm{d} \umu \ P(k,\umu) L_l (\umu). 
\end{equation}
The AP distortions will distort the power spectrum,
\begin{equation}
    P(k,\umu) = P^\mathrm{t} \left ( k^\mathrm{t} \left [ k,\umu \right ],\umu^\mathrm{t}\left [k,\umu \right ] \right ),\label{eq:Pfulldistortion}
\end{equation}
and its multipoles.

To linear order, the measured multipoles can be related to true multipoles as
\begin{align}
    \nonumber
    P = \Pt &+ \frac{\upartial \Pt}{\upartial k}\left(\frac{\upartial k}{\upartial\aV}\delta\aV + \frac{\upartial k}{\upartial\alpha_{\epsilon}}\delta\alpha_{\epsilon}\right) + \\
    &+ \frac{\upartial \Pt}{\upartial \umu}\left(\frac{\upartial \umu}{\upartial\aV}\delta\aV + \frac{\upartial \umu}{\upartial\alpha_{\epsilon}}\delta\alpha_{\epsilon}\right).
\end{align}
Using Eqs~\eqref{eq:klinear}-\eqref{eq:umulinear} this becomes
\begin{equation}
\label{powerlog}
    P  = \Pt + \frac{\upartial\Pt}{\upartial \ln k} \delta\alpha_V + \frac{\upartial\Pt}{\upartial \ln k} L_2(\umu)\delta\alpha_{\epsilon} + \frac{\upartial\Pt}{\upartial \ln \umu} \frac{3}{2}\nu^2\delta\alpha_{\epsilon}.
\end{equation}

For the multipoles we get,
\begin{align}
\nonumber
 P_\ell &= \frac{2l+1}{2} \displaystyle\sum_{\ell'} \displaystyle\int \mathrm{d} \umu \left[ P^\mathrm{t}_{\ell'}L_{\ell'}(\umu) \frac{\upartial P^\mathrm{t}_{\ell '}}{\upartial \ln k}L_{\ell'}(\umu)\delta\alpha_V + \right. \\  
  &+ \left. \frac{\upartial P^\mathrm{t}_{\ell '}}{\upartial \ln k}L_{\ell'}(\umu)L_2(\umu)\delta\alpha_{\epsilon} + P^\mathrm{t}_{\ell'}  \frac{\upartial L_{\ell'}(\umu)}{\upartial \ln\umu} \frac{3}{2}\nu^2\delta\alpha_{\epsilon} \right] L_{\ell} (\umu),
\end{align}

which results in
\begin{align}
\nonumber
 P_\ell &= P^\mathrm{t}_\ell + \frac{\upartial P^\mathrm{t}_{\ell}}{\upartial \ln k}\delta\alpha_V + \displaystyle\sum_{l^{'}}  \frac{\upartial P^\mathrm{t}_{\ell'}}{\upartial \ln k} A_{l {l^{'}}}\delta\alpha_{\epsilon} + \\ 
 &+ \displaystyle\sum_{l^{'}}P_{l^{'}} ^\mathrm{t} B_{l {l^{'}}}\delta\alpha_{\epsilon}.\label{eq:Pexpansion}\\
 A_{ll'} &= \frac{2\ell+1}{2}\displaystyle\int \mathrm{d}\umu \ L_{\ell'}(\umu)L_\ell(\umu) L_2(\umu),\label{eq:Pcoeff1}\\
 B_{ll'} & = \displaystyle\int \mathrm{d} \umu \ \frac{\upartial L_{\ell'}(\umu)}{\upartial \ln\umu} L_{\ell} (\umu)\frac{3}{2}\nu^2 .\label{eq:Pcoeff2}
 \end{align}
The leading order term stretches each multipole by a factor of $k\delta\aV$, a constant fractional stretching by a factor of $\aV$ for each wavenumber. The second and the third terms mix different multipoles and only depend on $\delta\alpha_{\epsilon}$. The mixing coefficients can be computed analytically and are given in table~\ref{tab:all} for the first four multipoles. \citet{2008PhRvD..77l3540P} use the $\epsilon$ instead of $\alpha_\epsilon$ so our coefficients need to be multiplied by two to match their formulas since to linear order $\delta\epsilon = 2\delta\alpha_\epsilon$.

 \begin{table}
      \caption{All nonzero mixing coefficients for the power spectrum multipoles up to the fourth degree.}
     \label{tab:all}
     \begin{tabular}{|c|c|c|c|}
     \hline
          $\ell$ & $\ell'$ & $A_{\ell\ell'}$ & $B_{\ell\ell'}$  \\
          \hline
          0 & 2 & $\frac{1}{5}$ & $\frac{3}{5}$  \\
          \hline
          2 & 0 & $1$ & 0  \\
          \hline
          2 & 2 & $\frac{2}{7}$ & $\frac{3}{7}$  \\
          \hline
          2 & 4 & $\frac{2}{7}$& $\frac{10}{7}$  \\
          \hline
          4 & 2 & $\frac{18}{35}$ & $-\frac{36}{35}$ \\
          \hline
          4 & 4 & $\frac{20}{77}$ & $\frac{30}{77}$\\  
          \hline
     \end{tabular}
 \end{table}

\section{Distortions of the Bispectrum}
\label{sec:bispectrum}

Bispectrum is a function of five parameters and is usually expanded in spherical multipoles,
\begin{equation}
B(k_1,k_2,k_3,\umu,\phi) = \displaystyle\sum_{\ell m} B_{\ell m}(k_1,k_2,k_3)Y_{\ell m}(\theta_1,\phi),
\end{equation}
where $\theta_1 = \arccos(\umu_1)$.
The multipoles can be computed via
\begin{equation}
B_{\ell m}(k_1,k_2,k_3) = \displaystyle\int \mathrm{d}\umu_1\mathrm{d}\phi \ B(k_1,k_2,k_3,\umu_1,\phi)Y_{\ell m}(\theta_1,\phi) .
\end{equation}
The multipoles with nonzero $m$ have negligible amplitudes and we will ignore the second index for the rest of the paper.
The measured and the true bispectra are related by
\begin{equation}
    B(k_1,k_2,k_3,\umu_1,\phi) = B^\mathrm{t}(k_1^\mathrm{t},k_2^\mathrm{t},k_3^\mathrm{t},\umu_1^\mathrm{t},\phi^\mathrm{t}),
\end{equation}
where the right hand side is implicitly a function of measured parameters through Eqs.~\eqref{eq:kfull} and \eqref{eq:umufull} with the angles given by Eqs.~\eqref{eq:umu1}-\eqref{eq:umu3}. To linear order the distortions are given by
\begin{align}
    \nonumber
    B = B^\mathrm{t} & + \displaystyle\sum_{i=1,2,3}\frac{\upartial B^\mathrm{t}}{\upartial k_i}\left(\frac{\upartial k_i}{\upartial\aV}\delta\aV + \frac{\upartial k_i}{\upartial\alpha_{\epsilon}}\delta\alpha_{\epsilon}\right) +
    \\
    &+\frac{\upartial B^\mathrm{t}}{\upartial \umu_1}\left(\frac{\upartial \umu_1}{\upartial\aV}\delta\aV + \frac{\upartial \umu_1}{\upartial\alpha_{\epsilon}}\delta\alpha_{\epsilon}\right) + \\
    & + \frac{\upartial B^\mathrm{t}}{\upartial \phi}\left(\frac{\upartial \phi}{\upartial\aV}\delta\aV + \frac{\upartial \phi}{\upartial\alpha_{\epsilon}}\delta\alpha_{\epsilon}\right).\nonumber
\end{align}
We checked that $\phi$-derivatives have a subdominant effect and we will ignore them for the rest of the paper. Using Eqs.~\eqref{eq:klinear}-\eqref{eq:umulinear}, we get
\begin{multline}
    B = B^\mathrm{t} + \mathlarger{\mathlarger{\sum}}_{i=1,2,3}\frac{\upartial B^\mathrm{t}}{\upartial \ln{k_i}} \delta \alpha_{V} + \mathlarger{\mathlarger{\sum}}_{i=1,2,3}\frac{\upartial B^\mathrm{t}}{\upartial \ln{k_i}}L_2(\umu_i)\delta \alpha_{\epsilon}
    + \\
    + \frac{\upartial B^\mathrm{t}}{\upartial \ln{\umu_1}}\frac{3}{2}\nu_1^2 \delta\alpha_{\epsilon} .
\end{multline}
Expanding both sides of the equation into multipoles results in
\begin{multline}
   B_{\ell} = B^\mathrm{t}_{\ell} + \displaystyle\sum_{i=1}^{3} \frac{\upartial B^\mathrm{t}_{\ell}}{\upartial \ln k_i}\delta\alpha_V + \displaystyle\sum_{\ell',m',i=1}^{3} \frac{\upartial B^\mathrm{t}_{\ell' m'}}{\upartial \ln k_i}C^{(i)}_{\ell\ell'm'}\delta\alpha_{\epsilon}  + 
   \\
   +   \sum_{\ell'}\frac{\upartial B^\mathrm{t}_{\ell'}}{\upartial \ln \umu_1}D_{\ell\ell'}\delta\alpha_{\epsilon},\label{eq:Bexpansion}
\end{multline}
where,
\begin{align}
C^{(i)}_{\ell\ell'm'} &= \displaystyle\int \mathrm{d}\umu_1\mathrm{d}\phi\ L_2(\umu_i)Y_{\ell 0}(\umu_1)Y_{\ell'm'}(\umu_1)\label{eq:Bcoeff1}\\
D_{\ell\ell'} &= \frac{3}{2}\displaystyle\int \mathrm{d}\umu_1\mathrm{d}\phi\ \frac{\upartial Y_{\ell' m'}(\umu_1)}{\upartial \ln\umu_1}\nu_1^2Y_{\ell0}(\umu_1)\label{eq:Bcoeff2}
\end{align}

The three leading terms tell us that, similarly to the power spectrum, all three wavevectors stretch with a relative scaling factor of $\delta\aV$. The remaining terms mix the multipoles (including non-zero $m$ multipoles). They can be computed analytically and their values are given in tables~\ref{tab:Ccoeff} and \ref{tab:Dcoeff}. The $C$-coefficients are symmetric with respect to $\ell$ to $\ell'$ interchange, while the $D$-coefficients are not. These coefficients will have to be multiplied by a factor of two when using $\epsilon$ as the anisotropic warping parameter.

\begin{table}
     \caption{All nonzero coefficients $C^{(i)}_{\ell\ell'm'}$ up to $\ell = 4$.}
     \label{tab:Ccoeff}
     \begin{tabular}{|c|c|c|c|c|c|}
     \hline
          $\ell$ & $\ell'$  & $m'$ & $C^{(1)}_{\ell\ell'm'}$ &$C^{(2)}_{\ell\ell'm'}$ & $C^{(3)}_{\ell\ell'm'}$  \\
  \hline
   
   0 & 2 & 0 & $\frac{1}{\sqrt{5}}$ &  $\frac{1}{\sqrt{5}} L_2(\umu_{12}) $ & $\frac{1}{\sqrt{5}}L(\umu_{31})$\\
  \hline
   0 & 2 & 2 & 0 & $\frac{1}{2}\sqrt{\frac{3}{10}}\nu_{12}^2$ & $\frac{1}{2}\sqrt{\frac{3}{10}}\nu_{31}^2$\\
  \hline
    2 & 0 & 0 & $\frac{1}{\sqrt{5}}$ & $\frac{1}{2\sqrt{5}}L_2(\umu_{12})$ & $\frac{1}{2\sqrt{5}}L_2(\umu_{31})$ \\
  \hline
    2 & 2 & 0 & $\frac{2}{7}$ & $\frac{2}{7}L_2(\umu_{12})$ & $\frac{2}{7}L_2(\umu_{31})$ \\
  \hline
    2 & 2 & 2 & 0 & -$\frac{1}{7}\sqrt{\frac{3}{2}}\nu_{12}^2$ & -$\frac{1}{7}\sqrt{\frac{3}{2}}\nu_{31}^2$\\
  \hline
  2 & 4 & 0 & $\frac{6}{7\sqrt{5}}$ & $\frac{6}{7\sqrt{5}}L_2(\umu_{12})$ & $\frac{6}{7\sqrt{5}}L_2(\umu_{31})$\\
  \hline
   2 & 4 & 2 & 0 & $\frac{3}{14\sqrt{2}} \nu_{12}^2$ & $\frac{3}{14\sqrt{2}} \nu_{31}^2$ \\
  \hline
  4 & 2 & 0 & $\frac{6}{7\sqrt{5}}$ & $\frac{6}{7\sqrt{5}}L_2(\umu_{12})$ & $\frac{6}{7\sqrt{5}} L_2(\umu_{31})$  \\
  \hline
  4 & 2 & 2 & 0 & $\frac{1}{14}\sqrt{\frac{3}{10}} \nu_{12}^2$ & $\frac{1}{14}\sqrt{\frac{3}{10}} \nu_{31}^2 $ \\
  \hline
4 & 4 & 0 & $\frac{20}{77}$ & $\frac{20}{77}L_2(\umu_{12})$ & $\frac{20}{77}L_2(\umu_{31})$ \\
  \hline
  4 & 4 & 2 & 0 & $-\frac{9}{77}\sqrt{\frac{5}{2}} \nu_{12}^2$ & $-\frac{9}{77}\sqrt{\frac{5}{2}} \nu_{31}^2$  \\
  \hline
  4 & 6 & 0 & $\frac{15}{11\sqrt13}$ & $\frac{15}{11\sqrt{13}}L_2(\umu_{12})$ & $\frac{15}{11\sqrt{13}}L_2(\umu_{31})$  \\
  \hline
    4 & 6 & 2 & 0 & $\frac{1}{22}\sqrt{\frac{105}{13}}\nu_{12}^2$ & $\frac{1}{22}\sqrt{\frac{105}{13}}\nu_{31}^2$   \\
  \hline
  
    \end{tabular}
 \end{table} 

 \begin{table}
      \caption{All nonzero $D_{\ell\ell'}$ coefficients up to $\ell=4$.}
     \label{tab:Dcoeff}
     \begin{tabular}{|c|c|c|}
     \hline
          $\ell$ & $\ell'$ & $D_{\ell\ell'}$\\
          \hline
          0 & 2 & $\frac{3}{\sqrt{5}}$\\
          \hline
         2 & 2 & $\frac{3}{7}$\\
         \hline
         2 & 4 & $\frac{6}{7\sqrt{5}}$\\
         \hline
         4 & 2 & $-\frac{12}{7\sqrt{5}}$\\
         \hline
          4 & 4 & $\frac{30}{77}$\\
          \hline
         4 & 6 & $\frac{105}{11\sqrt{13}}$\\ 
         \hline
     \end{tabular}
 \end{table} 

\section {Testing Linear Approximation}
\label{sec:validation}

\begin{figure}
    \includegraphics[width=\columnwidth]{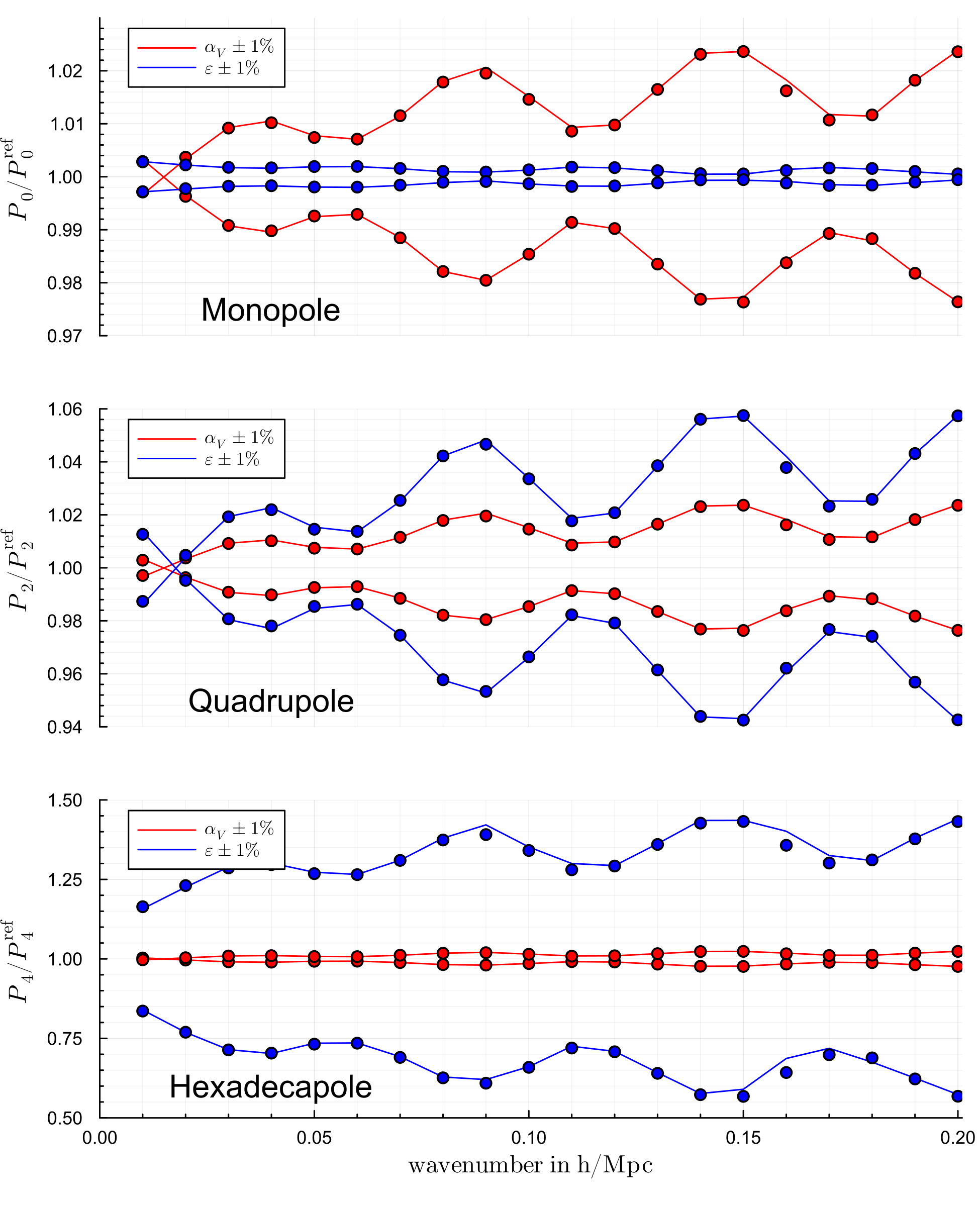}
    \caption{The ratio of distorted power spectrum multipoles to the fiducial (undistorted) ones. The points represent the results from the exact formula, while the lines represent the computations based on the linear approximation. The panels show the monopole, the quadrupole, and the hexadecapole of the power spectrum from top to bottom respectively. Red data points and lines correspond to a one per cent offset in $\aV$, while the blue data points and lines correspond to a one per cent offset in $\alpha_{\epsilon}$.}
    \label{fig:Pap}
\end{figure}

To test the validity of Eqs.~\eqref{eq:Pexpansion}-\eqref{eq:Pcoeff2} and \eqref{eq:Bexpansion}-\eqref{eq:Bcoeff2} we compute power spectrum and bispectrum in linear theory using $\Lambda$CDM model (see Appendix~\ref{app:PandBcomputation}) with best-fit parameters as constrained by the \texttt{Planck} satellite mission  \citep{2020A&A...641A...6P}. We first compute the power spectrum by distorting the wavenumbers following the exact formula in Eq.~\eqref{eq:Pfulldistortion}. We then compute the linear approximation of Eq.~\eqref{eq:Pexpansion}. The ratio of these two computations to the undistorted power spectrum is shown on figure~\ref{fig:Pap}.

The figure shows the ratios for the first three even multipoles of the power spectrum in the wavelength range $0 < k < 0.2h^{-1}$ Mpc. $\aV$ has a significantly larger effect on the monopole than the $\alpha_{\epsilon}$ for the same level of distortion. A one per cent deviation in $\aV$ results in about a two per cent deviation in the monopole amplitude, with a clearly visible BAO signature. The linear approximation is in excellent agreement with the exact result. $\aV$-distortions are smaller than $\alpha_{\epsilon}$-distortions in the quadrupole, but they are not negligible. A one per cent distortion in $\alpha_{\epsilon}$ results in a six per cent offset in the original amplitude of the quadrupole, while a similar distortion in the $\aV$ results in a twp per cent offset. Both distortions have clear BAO signals imprinted in them. $\alpha_{\epsilon}$-distortions dominate the quadrupole. A one per cent offset in $\alpha_{\epsilon}$ results in up to 30 per cent offset in the quadrupole amplitude. the linear approximation works exceedingly well for all three multipoles. 

\begin{figure}
    \includegraphics[width=\columnwidth]{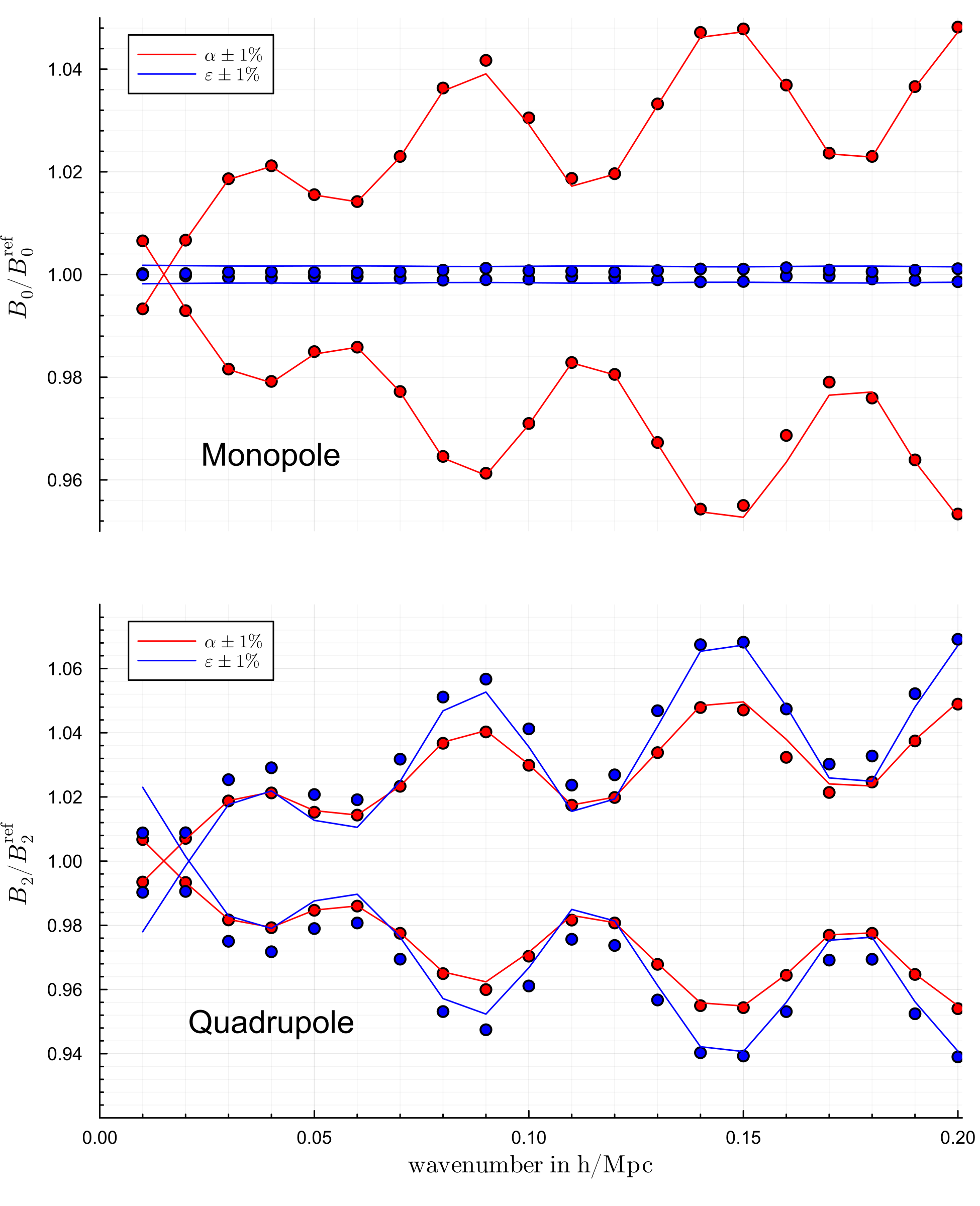}
    \caption{The ratio of distorted bispectrum multipoles to the fiducial (undistorted) ones. The points represent the results from the exact formula, while the lines represent the computations based on the linear approximation. The panels show the monopole and  the quadrupole of the bispectrum from top to bottom respectively. Red data points and lines correspond to a one per cent offset in $\aV$, while the blue data points and lines correspond to a one per cent offset in $\alpha_{\epsilon}$.}
    \label{fig:Bap}
\end{figure}

\begin{figure}
    \includegraphics[width=0.9\columnwidth]{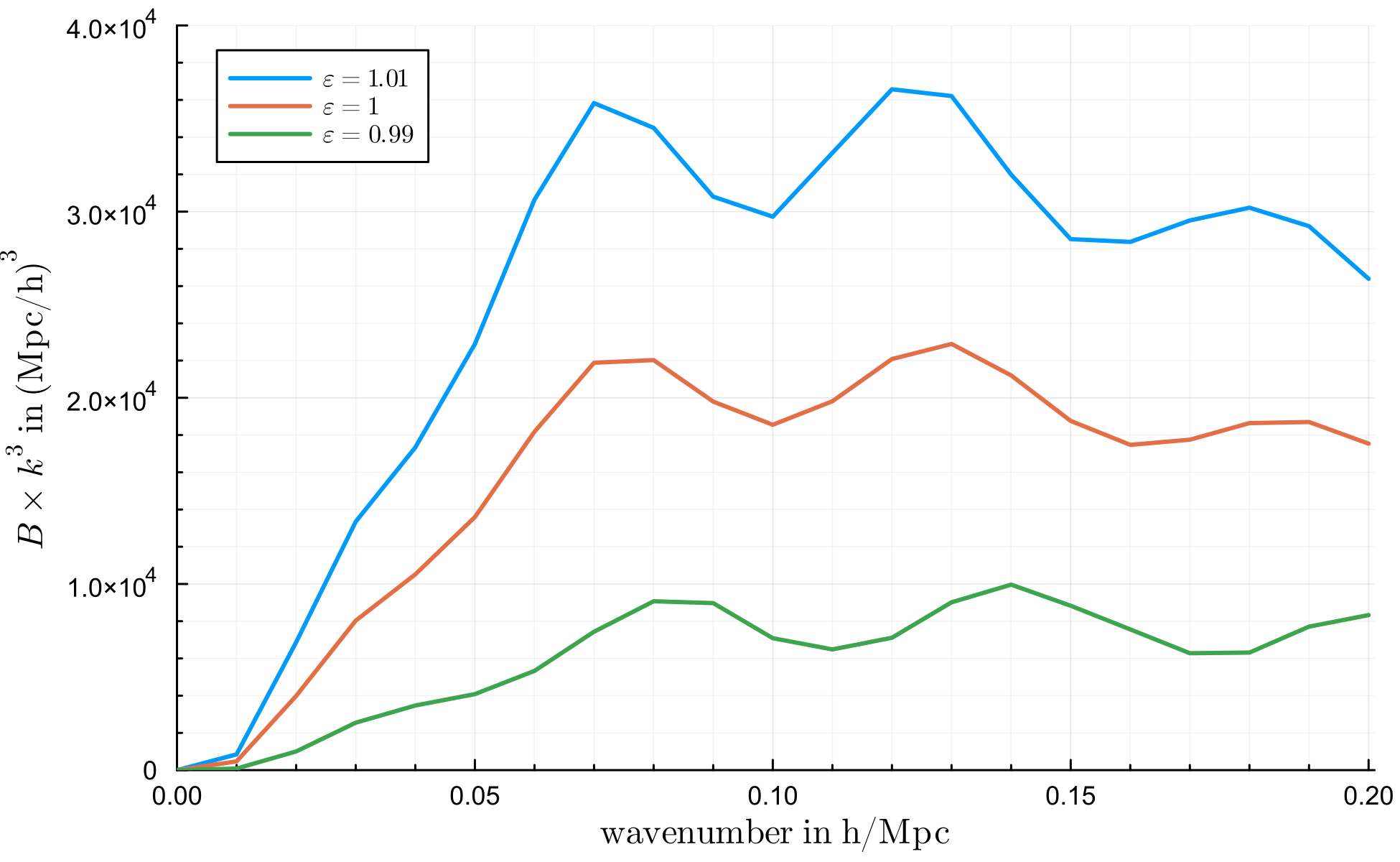}
    \caption{Response of the Bispectrum hexadecapole (equilateral configuration) to a one percent offset in $\alpha_{\epsilon}$.}
    \label{fig:hex}
\end{figure}

Figure~\ref{fig:Bap} shows a similar ratio for the bispectrum. The bispectrum monopole is completely dominated by the $\aV$-distortions. one per cent offset in $\aV$ results in a four per cent offset in the monopole amplitude, and the BAO feature is clearly pronounced. $\alpha_{\epsilon}$ and $\aV$ distortions have comparable effects on the bispectrum quadrupole, both resulting in a four per cent offset of the amplitude for one per cent distortion. The linear approximation works extremely well for the bispectrum monopole. We start seeing some deviation between the linear prediction and the exact result for the bispectrum quadrupole at low wavevectors. This deviation is mostly in amplitude and preserves the position of the BAO peaks. The bispectrum hexadecapole is extremely strongly affected by the $\alpha_{\epsilon}$-distortions and the linear expansion fails even for small distortions.
Figure~\ref{fig:hex} shows the effect of a one per cent distortion in the bispectrum hexadecapole. The amplitude of hexadecapole changes by a factor of two even for such a small distortion.

\section{Conclusions}

\label{sec:conclusions}

We derived explicit analytic expressions for the mixing coefficients of the bispectrum multipoles to the linear order in Alcock-Patczynski distortions. They are presented in Tables~\ref{tab:Ccoeff} and \ref{tab:Dcoeff} for the multipoles up to $\ell = 4$. The coefficients are derived for the natural definition of the bispectrum multipoles but can easily be extended to other parametrizations and definitions \citep[e.g. the ones proposed in][]{2019MNRAS.484L..29G, 2019MNRAS.484..364S}. We find that the linear order mixing coefficients work extremely well for all power spectrum multipoles (see figure~\ref{fig:Pap}). For the bispectrum monopole the leading order mixing model works very well. The isotropic stretching (non-zero $\aV$) is modeled very accurately by the linear expression for all multipoles. The linear mixing model is not so accurate for the anisotropic stretching (non-zero $\alpha_{\epsilon}$) (see figure~\ref{fig:Bap}). Predictions for the bispectrum quadrupole deviate by a fraction of a per cent for $\alpha_{\epsilon} \sim 1\%$. Even small values of $\alpha_{\epsilon} \sim 1\%$ induce large changes in the amplitude and shape of the bispectrum hexadecapole and the linear approximation obviously fails. Our findings will be helpful for future analysis of galaxy bispectrum, especially for the works that aim to measure the BAO feature in the galaxy bispectrum. The computational cost of modeling AP distortions of the bispectrum monopole and quadrupole can be significantly reduced by using the linear mixing model around a fiducial value as opposed to directly computing the distorted monopoles. The linear mixture model does not seem to be sufficiently accurate for the bispectrum quadrupole and the computationally more expansive, direct computation may be unavoidable. 

\section*{acknowledgements}
The authors are grateful for the support from the US Department of Energy via grants DE-SC0021165 and DE-SC0011840. LS is grateful for support from the NASA ROSES grant 12-EUCLID12-0004 and the Shota Rustaveli National Science Foundation of Georgia grants FR
19-498.

\section*{data availability}

The code used to make figures in this manuscript is available at \url{https://github.com/ladosamushia/BispectrumAP}.

\bibliographystyle{mnras}
\bibliography{main}
\appendix

\newpage
\onecolumn
\section{Wavevector Projections}
\label{app:vectorprojections}

Our goal is to derive expressions for the cosine of the angle with respect to the z-axis for an arbitrary triangle shown on Figure~\ref{fig:triangle}. 

We will start with a triangle in the y-z plane with a  $k_1$ vector along the positive y-axis. To get the triangle on Fig.~\ref{fig:triangle} we have to rotate the initial triangle by angle $\phi$ counterclockwise with respect to the y-axis and by the angle $\upi/2 - \theta_1$ counterclockwise with respect to the x-axis.

Matrices corresponding to these rotations are
\begin{equation}
R_y = 
    \begin{pmatrix}
     \cos(\phi) & 0 & \sin(\phi) \\
    0 & 1 & 0 \\
    -\sin(\phi) & 0 & \cos(\phi) 
    \end{pmatrix},
\end{equation}
and
\begin{equation}
R_x = 
    \begin{pmatrix}
     1 & 0 & 0 \\
    0 & \sin(\theta) & -\cos(\theta) \\
    0 & \cos(\theta) & \sin(\theta) 
    \end{pmatrix}.
\end{equation}

The initial positions of wavevectors are
\begin{align}
\mathbf{k}_{1,\mathrm{ini}} &= (0, k_1, 0)^\mathrm{T},\\
\mathbf{k}_{2,\mathrm{ini}} &= (0,-k_2\umu_{12},k_2\nu_{12})^\mathrm{T},\\
\mathbf{k}_{3,\mathrm{ini}} &= (0,-k_3\umu_{31},-k_3\nu_{31})^\mathrm{T},
\end{align}
where $\theta_{ij}$ is the angle between vectors $k_i$ and $k_j$.

After the rotation we get
\begin{align}
\mathbf{k}_1 &= R_xR_y\mathbf{k}_{1,\mathrm{ini}} = (0, k_1\nu_\theta, k_1\umu_\theta)^\mathrm{T},\\
\nonumber
\mathbf{k}_2 &= R_xR_y\mathbf{k}_{2,\mathrm{ini}} = (k_2\nu_{12}\nu_\phi,-k_2\umu_\theta\umu_\phi\nu_{12}-k_2\umu_{12}\nu_\theta,-k_2\umu_{12}\umu_\theta + k_2\umu_\phi\nu_{12}\nu_\theta)^\mathrm{T},\\
\nonumber
\mathbf{k}_3 &= R_xR_y\mathbf{k}_{3,\mathrm{ini}} = (k_3\nu_{31}\nu_\phi, k_3\umu_\theta\umu_\phi\nu_{31}-k_2\umu_{31}\nu_\theta,-k_3\umu_{31}\umu_\theta - k_3\umu_\phi\nu_{31}\nu_\theta)^\mathrm{T},
\end{align}
and the cosine of the angle with respect to the line of site is
\begin{align}
    \umu_1 &= \umu_\theta,\\
    \umu_2 &= -\umu_{12}\umu_\theta + \umu_\phi\nu_{12}\nu_\theta,\\
    \umu_3 &= -\umu_{31}\umu_\theta - \umu_\phi\nu_{31}\nu_\theta.
\end{align}

\section{Bispectrum projections from Geometry}

The results equivalent to the ones in the appendix~\ref{app:vectorprojections} can be derived directly by projecting the triangle on Figure~\ref{fig:triangle} by using the identities of Euclidean geometry. The angle and the magnitude of $\mathbf{k}_1$ are trivial. For the other two wavevectors we get the following expressions.
\begin{equation} 
    \label{k3}
    k_2^t = \sqrt{k_2^2 + ( \alpha_{||}-1)^2 k_1 \umu} - k_3 \sqrt{1-\frac{r^2 \nu^2_\phi}{k_3^2}} \cos (\alpha_{\epsilon}^\phi + \theta)^2 + ( \alpha_\perp-1)^2 ( k_1 \nu - k_3 \sqrt{1-\frac{r^2 \nu^2_\phi}{k_3^2}} \sin( \alpha_{\epsilon}^\phi + \theta )^2 +  \frac{r^2 \nu_\phi^2}{k_3} \left( \alpha_\perp-1 \right )^2
\end{equation}
\begin{equation}
    k_3^t = k_3 \cdot  \sqrt{\left ( 1- \frac{r^2 \sin^2{\phi}}{k_3^2} \right ) \left [ \cos^2{\left ( \alpha_{\epsilon}^\phi + \theta \right )}  \alpha_{||}^2 + \sin^2{\left ( \alpha_{\epsilon}^\phi + \theta \right )} \alpha_{\perp}^2 \right ] +  \frac{r^2 \sin^2{\phi}}{k_3^2} \alpha_{\perp}^2},
    \end{equation}
where
\begin{equation}
    {\alpha_{\epsilon}}^{\phi} = \arctan \left ( {\frac{r \cos{\phi}}{d}} \right )
\end{equation}
and
\begin{equation}
    {\xi}^{\phi} = \arctan \left ( {\frac{r \cos{\phi}}{k_1 - d}} \right )
\end{equation}

To linear order in small distortions these becomes
\begin{multline} \label{k3lin}
    k_2^t = k_2 + \delta \alpha_{||} \left [ k_1 \umu - k_3 \sqrt{1-\frac{r^2 \sin^2{\phi}}{{k_3}^2}} \cdot \cos{\left ( \alpha_{\epsilon}^\phi + \theta \right )}  \right ] + 
    \\
    + \delta \alpha_{\perp} \left [ k_1 \nu - k_3 \sqrt{1-\frac{r^2 \sin^2{\phi}}{{k_3}^2}} \cdot \sin{\left ( \alpha_{\epsilon}^\phi + \theta \right )}  \right ] +\delta \alpha_{\perp} \cdot \frac{{r^2 \sin^2{\phi}}}{k_2} 
\end{multline}
\begin{equation}
    k_3^t  = k_3 \left ( 1- \frac{r^2 \sin^2{\phi}}{k_3^2} \right ) \left [ \cos^2{\left ( \alpha_{\epsilon}^\phi + \theta \right )} \delta \alpha_{||} + \sin^2{\left ( \alpha_{\epsilon}^\phi + \theta \right )} \delta \alpha_{\perp} \right ] +
    k_3 \cdot \frac{r^2 \sin^2{\phi}}{k_3^2} \delta \alpha_{\perp}+ k_3 
\end{equation}

Distortions in $\phi$ are given by
\begin{equation}
    \phi^t = asin \left ( \frac{r}{r^t} sin\phi \left ( 1 + \delta \alpha_{\perp} \right ) \right ),
\end{equation}
where 
\begin{equation}
    r=\sqrt{{k_3}^2 - {d}^2},
\end{equation}
and
\begin{equation}
    d=\frac{{k_1}^2 + {k_3}^2-{k_2}^2}{2k_1}
\end{equation}

The distortions in $r$ to linear order are
\begin{equation}
    \frac{r}{r^t}=1+ \delta k_1 - \frac{1}{r^2 k_1^2} \left [ \sum_{i,j}^{i \neq j} k_i^2 k_j^2 \cdot \delta k_j - \sum_{i=1}^3 k_i^4 \cdot \delta k_i \right]
\end{equation}
which results in the following linearized expression for the $\phi$ -
\begin{equation}
    \phi^\mathrm{t} = \phi + \tan{\phi}\left ( \delta \alpha_{\perp} + \delta k_1 - \frac{1}{r^2 k_1^2} \left [ \sum_{i,j}^{i \neq j} k_i^2 k_j^2 \cdot \delta k_j - \sum_{i=1}^3 k_i^4 \cdot \delta k_i \right] \right ).
\end{equation}

For equilateral configurations, with $k_1=k_2=k_3=K$, the equations above simplify further to

\begin{equation}
k_1^\mathrm{t} = K(1 + \umu^2\delta\alpha_\parallel+\nu^2\delta\alpha_\perp)
\end{equation}

\begin{multline} 
    k_2^t = K + \delta \alpha_{||} \left [ K \umu - K \sqrt{1-\frac{3 \sin^2{\phi}}{4}} \cdot \left ( \frac{\umu}{\sqrt{1 + 3 \cos^2 \phi}} - \frac{\sqrt{3} \cos \phi \cdot \nu}{\sqrt{1+3 \cos^2 \phi}}  \right )  \right ] 
    \\
    + \delta \alpha_{\perp} \left [ K \nu - K \sqrt{1-\frac{3 \sin^2{\phi}}{4}} \cdot \left ( \frac{\sqrt{3} \cos \phi \cdot \umu}{\sqrt{1+3 \cos^2 \phi}} + \frac{\nu}{\sqrt{1 + 3 \cos^2 \phi}} \right )  \right ] +\delta \alpha_{\perp} \cdot K \frac{{\sin^2{\phi}}}{4}
\end{multline}

\begin{align}
k_3^\mathrm{t} &= K + K\left ( 1- \frac{3 \sin^2{\phi}}{4} \right ) \left [ \left ( \frac{\umu}{\sqrt{1 + 3 \cos^2 \phi}} - \frac{\sqrt{3} \cos \phi \cdot \nu}{\sqrt{1+3 \cos^2 \phi}} \right )^2 \delta \alpha_{||} + \left ( \frac{\sqrt{3} \cos \phi \cdot \umu}{\sqrt{1+3 \cos^2 \phi}} + \frac{\nu}{\sqrt{1 + 3 \cos^2 \phi}} \right )^2 \delta \alpha_{\perp} \right ] \\ \nonumber
&+ K \cdot \frac{\sin^2{\phi}}{4} \delta \alpha_{\perp}
\end{align}

\begin{equation}
    \phi^t = \arctan \left ( \sin{\phi} \left ( 1 + \delta \alpha_{\perp} + \delta k_1 -\frac{2}{3} (\delta k_1 + \delta k_2 +\delta k_3) \right ) \right )
\end{equation}

\section{Computing Model Power Spectrum And Bispectrum}
\label{app:PandBcomputation}

We use linear perturbation theory to compute the model power spectrum and bispectrum for our validation tests presented on Figures~\ref{fig:Pap} and \ref{fig:Bap} \citep{ 2000ApJ...544..597S,2017MNRAS.467..928G}. 

We first compute the linear power spectrum $P_\mathrm{lin}$ using \texttt{CAMB}\footnote{\url{https://lambda.gsfc.nasa.gov/toolbox/camb_online.html}} software \citep{Lewis:2002ah}. The galaxy power spectrum is then computed as
\begin{equation}
    P({k}) = \left ( b_1 + f \umu^2 \right ) P_\mathrm{lin}(k),
\end{equation}
where $b_1$ and $f$ are free parameters. We compute the bispectrum using
\begin{equation}
B \left ( k_1,k_2,k_3 \right ) = 2 Z_1 (\umu_1) Z_1 (\umu_2) Z_2(\mathbf{k}_1,\mathbf{k}_2) P_\mathrm{lin}(k_1) P_\mathrm{lin}(k_2) + \textrm{cyclic}\ \textrm{terms}
\end{equation}
where
\begin{equation}
    Z_1 (\umu) = \left ( b_1 + f \umu^2 \right )
\end{equation}
\begin{equation}
    Z_2(\mathbf{k}_1, \mathbf{k}_2) = \Biggl\{ \frac{b_2}{2} + b_1 F_2 (\mathbf{k}_1,\mathbf{k}_2) + f \umu_3^2 G_2(\mathbf{k}_1,\mathbf{k}_2) - \frac{f \umu_3 k_3}{2} \left [ \frac{\umu_1}{k_1} (b_1 + f \umu_2^2) + \frac{\umu_2}{k_2} (b_1 + f \umu_1^2) \right ] \Biggl\}
\end{equation}
\begin{equation}
   F_2(k_1,k_2) = \frac{5}{7} + \frac{\mathbf{k}_1  \mathbf{k}_2}{2k_1 k_2} \left ( \frac{k_1}{k_2} + \frac{k_2}{k_1} \right ) + \frac{2}{7} \left ( \frac{\mathbf{k}_1  \mathbf{k}_2}{2k_1 k_2}  \right )^2
\end{equation}
\begin{equation}
   G_2(k_1,k_2) = \frac{3}{7} + \frac{\mathbf{k}_1 \mathbf{k}_2}{2k_1 k_2} \left ( \frac{k_1}{k_2} + \frac{k_2}{k_1} \right ) + \frac{4}{7} \left ( \frac{\mathbf{k}_1 \mathbf{k}_2}{2k_1 k_2}  \right )^2.
\end{equation}
Cyclic terms are computed by permuting the indeces (replacing 1 and 2, by 2 and 3, and then by 3 and 1 in the above equations). 


\label{lastpage}
\end{document}